\documentclass[%
]{algotel}
\usepackage[latin1]{inputenc}
\usepackage[francais]{babel}
\usepackage{float}
\usepackage{amsmath}
\usepackage{graphicx}
\usepackage{amssymb}

\usepackage{theorem}

\newtheorem{theoreme}{Th\'eor\`eme}
\newtheorem{definition}{D\'efinition}
\newtheorem{lemme}{Lemme}

\author{Alain Cournier\addressmark{1}, Swan Dubois\addressmark{2} et Vincent Villain\addressmark{1}}

\title{Une CNS pour l'acheminement de messages instantan\'ement stabilisant}

\address{
\addressmark{1}Laboratoire MIS, Universit\'e de Picardie Jules Verne, 33 rue Saint Leu, 80039 Amiens Cedex 1 (France)\\
\{alain.cournier,vincent.villain\}@u-picardie.fr\\
\addressmark{2}LIP6 - UMR 7606/INRIA Rocquencourt, \'Equipe-projet REGAL, Universit\'e Pierre et Marie Curie - Paris 6\\
104 Avenue du Président Kennedy, 75016 Paris (France), swan.dubois@lip6.fr
}


\begin{document}
\maketitle

\begin{abstract} 
Un algorithme \emph{instantan\'ement stabilisant} assure qu'il se comporte toujours conform\'ement \`a ses sp\'ecifications en 
partant d'une configuration initiale quelconque. Dans cet article, nous nous int\'eressons au probl\`eme de l'acheminement de 
messages dans un r\'eseau dot\'e d'une commutation de messages. Nous devons g\'erer les ressources du r\'eseau de mani\`ere \`a 
pouvoir d\'elivrer des messages \`a tout processeur du r\'eseau. Dans ce but, nous utilisons l'information fournie par un 
algorithme de routage. Mais, en raison du contexte de la stabilisation, cette information peut \^etre initialement incorrecte. 
C'est pourquoi l'existence d'algorithmes instantan\'ement stabilisants pour cette t\^ache (d\'emontr\'ee dans \cite{CDV09a}) 
implique que nous pouvons demander au syst\`eme de commencer \`a acheminer des messages m\^eme si les tables de routage sont 
initialement corrompues. Dans cet article, nous g\'en\'eralisons le r\'esultat pr\'ec\'edent en donnant une condition 
n\'ecessaire et suffisante pour r\'esoudre ce probl\`eme de mani\`ere instantan\'ement stabilisante.
\end{abstract}

\section{Introduction} 
\label{sec:introduction}

De nombreux concepts de tol\'erance aux pannes ont \'et\'e introduits en syst\`emes distribu\'es. Par exemple, l'auto-stabilisation 
(\cite{D74}) assure que le syst\`eme, ind\'ependament de son \'etat initial, retrouve un comportement r\'epondant \`a ses 
sp\'ecifications \emph{en un temps fini} sans intervention externe. L'\'etat initial quelconque peut permettre de mod\'eliser l'effet 
de fautes transitoires sur le r\'eseau. Un autre concept, la stabilisation instantan\'ee (\cite{BDPV07}), garantit que le syst\`eme, 
ind\'ependament de son \'etat initial, a \emph{toujours} un comportement r\'epondant \`a ses sp\'ecifications.
Dans un syst\`eme distribu\'e, il est classiquement suppos\'e que tout processeur ne peut communiquer directement qu'avec ses voisins. 
Pourtant, ce processeur peut avoir besoin de communiquer avec tout processeur du r\'eseau. Dans ce but, il y a en r\'ealit\'e deux 
probl\`emes \`a r\'esoudre : la d\'etermination du chemin \`a suivre par les messages (probl\`eme du routage) et la gestion des 
ressources du r\'eseau r\'eserv\'ees au transport des messages (probl\`eme de l'acheminement). Ces ressources sont des espaces m\'emoire
(appel\'es buffers) utilis\'es pour stocker temporairement les messages durant leur acheminement. Ces deux probl\`emes ont \'et\'e 
largement \'etudi\'es (\emph{cf.} \cite{T01,CDV09a} pour les r\'ef\'erences).
Nous disposons de nombreuses solutions stabilisantes pour le premier probl\`eme. C'est pourquoi dans cet article, nous nous int\'eressons 
aux solutions stabilisantes pour le deuxi\`eme. \cite{CDV09a} d\'emontre qu'il existe des algorithmes d'acheminement instantan\'ement 
stabilisants (\`a condition qu'un algorithme de calcul de table de routage auto-stabilisant s'ex\'ecute simultan\'ement). Cela implique 
que nous pouvons demander au syst\`eme de commencer \`a acheminer des messages m\^eme si les tables de routage sont initialement corrompues.
Dans la suite, nous appellerons message valide tout message qui a \'et\'e g\'en\'er\'e par un processeur (par cons\'equent, 
un mesage invalide est un message pr\'esent dans la situation initiale). Nous pouvons alors sp\'ecifier le probl\`eme de 
l'acheminement de la façon suivante :
\textbf{(1)} Tout message peut \^etre \'emis en un temps fini et
\textbf{(2)} Tout message valide sera d\'elivr\'e \`a son destinataire en un temps fini.\\
L'objectif de cet article est de pr\'esenter une condition n\'ecessaire et suffisante pour obtenir un algorithme d'acheminement 
instantan\'ement stabilisant. Il s'inspire d'un r\'esultat similaire obtenu dans un environnement sans fautes (\cite{MS78,TS81}). 
La suite de l'article est structur\'ee comme suit : dans un premier temps, nous pr\'esentons le r\'esultat obtenu dans \cite{MS78,TS81} 
pour un environnement sans fautes (section \ref{sec:theoreme}) puis nous donnons notre contribution dans la section \ref{sec:contribution}.

\section{Th\'eor\`eme dans un environnement sans fautes} 
\label{sec:theoreme}

Dans cette section, nous nous plaçons dans un syst\`eme distribu\'e qui ne peut pas subir de fautes. La configuration initiale est d\'efinie : 
les tables de routage sont correctes et il n'y a aucun message invalide dans les buffers.
Dans cet article, nous nous plaçons dans un r\'eseau \`a commutation de message (\emph{cf.} \cite{T01}). Chaque processeur dispose de $b\in\mathbb{N}$ 
buffers de taille suffisante pour contenir tout message. La m\'ethode de commutation est compos\'ee de trois types de mouvements :\\
\indent - \textbf{G\'en\'eration} : c'est la "cr\'eation" d'un nouveau message. Nous supposons que celle-ci est autoris\'ee d\`es qu'un 
buffer du processeur \'emetteur est libre.\\
\indent - \textbf{Transmission} : c'est la copie d'un message dans un buffer du processeur suivant sur le chemin calcul\'e par l'algorithme 
de routage. Nous supposons que ce mouvement est autoris\'e d\`es qu'un buffer sur le processeur en question est libre. En cons\'equence de 
ce mouvement, le buffer initial se lib\`ere en un temps fini.\\
\indent - \textbf{Consommation} : c'est le mouvement qui lib\`ere un buffer occupé par un message \`a destination du processeur sur lequel 
est situ\'e ce buffer. Le message est alors d\'elivr\'e \`a son destinataire. Nous supposons que ce mouvement est toujours autoris\'e.\\
Cependant, si nous n'effectuons aucun contr\^ole suppl\'ementaire sur les mouvements de message, le r\'eseau peut atteindre des situations 
inacceptables comme des interblocages. Il est alors impossible d'assurer les sp\'ecifications du probl\`eme. C'est pourquoi, il est n\'ecessaire 
de d\'efinir un algorithme (appel\'e contr\^oleur) qui autorise ou interdit dynamiquement (en fonction de l'occupation courante des buffers) 
certains mouvements. Si la r\'eponse \`a la question "Est-ce qu'un contr\^oleur $\mathcal{C}$ emp\^eche le r\'eseau d'atteindre un interblocage 
quelle que soit l'ex\'ecution issue de la configuration initiale ?" est affirmative, alors $\mathcal{C}$ est dit sans interblocage. En assurant 
l'absence de famine et de perte de messages, nous pouvons constater 
qu'un tel algorithme r\'epond au probl\`eme de l'acheminement de messages.
Nous allons pr\'esenter une classe de contr\^oleurs
sans interblocage basée sur le concept de \emph{graphe de buffers},
introduit par \cite{MS78}. Plus pr\'ecis\'ement, la structure choisie
est un DAG pour la raison suivante : un interblocage provient du fait qu'il existe
un circuit d'attente de lib\'eration de buffer. L'id\'ee de base est alors
de d\'efinir un DAG sur l'ensemble des buffers du r\'eseau de mani\`ere
\`a ce que les messages suivent les chemins de ce DAG. Ainsi, il ne
peut pas se former de circuit d'attente de lib\'eration de buffer.
Posons la notation suivante : $G=(V,E)$ est le graphe mod\'elisant
le r\'eseau ($V$ est l'ensemble des processeurs et $E$ l'ensemble
des liens de communications).

\vspace{-0.35cm}
\begin{definition}[Graphe de buffers]
Un graphe de buffers $BG=(\mathcal{B},BE)$ sur un graphe $G$ muni
d'un ensemble $\mathcal{B}$ de buffers et d'un ensemble $\mathcal{P}$
des plus courts chemins (induit par l'algorithme de routage) est d\'efini
de la mani\`ere suivante : 
\textbf{(1)} $BG$ est un graphe orient\'e,
\textbf{(2)} pour tout chemin $p\in\mathcal{P}$, il existe un chemin dans $BG$
dont la contraction est $p$\footnote{un chemin $c$ dans $BG$ est une suite de buffers $c=b_{1}...b_{t}$.
Notons alors $p_{i}$ le processeur sur lequel se situe $b_{i}$ pour
tout $i\in\{1,...,t\}$. La contraction de $c$ est alors $p_{1}...p_{t}$
en omettant les \'eventuelles r\'ep\'etitions.},
\textbf{(3)} pour chaque noeud $u$ de $G$ et pour chaque message $m$ possible,
il existe un buffer ad\'equat\footnote{un buffer $b$ est ad\'equat pour $m$ s'il existe un chemin de $b$ \`a
un buffer du destinataire de $m$ sur $BG$ dont
la contraction est dans $\mathcal{P}$.} de $BG$ not\'e $fb(m,u)$
sur $u$,
\textbf{(4)} pour chaque buffer $b$ de $\mathcal{B}$ situ\'e sur un processeur
$u$ et pour chaque message $m$ (non destin\'e à $u$) possible, il
existe un unique buffer ad\'equat, situ\'e sur $u$ ou sur un de ses voisins,
not\'e $nb(m,b)$ et
\textbf{(5)} $BE$ est l'ensemble des arcs $(b,nb(m,b))$ pour tout buffer $b\in\mathcal{B}$
et pour tout message $m$ (non destiné au processeur sur lequel est
$b$) possible.
\end{definition}
\vspace{-0.25cm}

La notation $fb(m,u)$, signifiant ''first buffer'', repr\'esente
le buffer dans lequel est plac\'e le message $m$ g\'en\'er\'e par le processeur
$u$. De m\^eme, la notation $nb(m,b)$, signifiant ''next buffer'',
repr\'esente le buffer dans lequel sera transmis le message $m$ qui
occupe le buffer $b$. Une fois un tel graphe de buffer d\'efini, il est possible de lui associer
un contr\^oleur selon la d\'efinition suivante :

\vspace{-0.35cm}
\begin{definition}[Contr\^oleur associ\'e \`a un graphe de buffers]
Etant donn\'e un graphe de buffers $BG=(\mathcal{B},$ $BE)$ sur un graphe
$G$ muni d'un ensemble $\mathcal{B}$ de buffers et d'un ensemble
$\mathcal{P}$ des plus courts chemins (induit par l'algorithme de
routage), nous d\'efinissons le contr\^oleur $\mathcal{C}_{BG}$ de la mani\`ere
suivante :\\
\indent - La g\'en\'eration d'un message $m$ sur un noeud $u$ est autoris\'e si
et seulement si $fb(m,u)$ est libre. $m$ est alors plac\'e dans ce
buffer.\\
\indent -  La transmission d'un message $m$ contenu dans le buffer $b$ est
autoris\'ee si et seulement si $nb(m,b)$ est libre, $m$ est alors
plac\'e dans ce buffer et $b$ est lib\'er\'e.
\end{definition}
\vspace{-0.25cm}

Nous sommes à pr\'esent en mesure de donner le r\'esultat fondamental
suivant (\cite{MS78,TS81}) :

\vspace{-0.35cm}
\begin{theoreme}
Soit un graphe de buffers $BG=(\mathcal{B},BE)$ sur un graphe $G$
muni d'un ensemble $\mathcal{B}$ de buffers et d'un ensemble $\mathcal{P}$
des plus courts chemins. Soit
le contr\^oleur $\mathcal{C}_{BG}$ associ\'e à $BG$ conform\'ement \`a la d\'efinition
2. Le contr\^oleur $\mathcal{C}_{BG}$ est sans interblocage si et seulement si
$BG$ est un DAG.
\end{theoreme}
\vspace{-0.25cm}

Le lecteur pourra trouver des exemples de tels contr\^oleurs dans \cite{T01}.

\section{Contribution} 
\label{sec:contribution}

Dans cette section, nous nous plaçons dans un syst\`eme distribu\'e sujet \`a des fautes transitoires. La configuration initiale est donc quelconque : 
les tables de routage sont incorrectes et il y a des messages invalides dans les buffers. Nous reprenons les notations et d\'efinitions introduites dans la section 2.
Soit $\mathcal{C}$ un contr\^oleur r\'epondant \`a la d\'efiniton 2 pour un graphe de buffers $BG$.
Nous supposons que, pour tout message $m$ occupant un buffer
$b$, le buffer $nb(m,b)$ est calcul\'e au moment de la transmission
du message en fonction des tables de routage \`a cet instant.
Nous supposons exister un algorithme $\mathcal{A}$ de calcul de table de routage auto-stabilisant et silencieux. 
Cet algorithme s'ex\'ecute de mani\`ere simultan\'ee mais prioritaire par rapport \`a $\mathcal{C}$ 
(noter que cet algorithme ne n\'ecessite que des communications entre voisins, il n'est donc pas utile qu'il utilise les buffers de communication distante g\'er\'es par $\mathcal{C}$).
Nous posons les notations suivantes : $FB$ et $NB$ repr\'esentent respectivement l'ensemble des buffers $fb(m,u)$ pour tout processeur $u$ et pour tout message possible $m$ associ\'es à $BG$ et l'ensemble des buffers $nb(m,b)$ pour tout buffer $b$ et pour tout message $m$ (non destin\'e au processseur sur lequel se situe $b$) associ\'es à $BG$.
Nous donnons \`a pr\'esent notre th\'eor\`eme :

\vspace{-0.35cm}
\begin{theoreme}[CNS pour obtenir un acheminement de messages instantan\'ement stabilisant]
$\mathcal{C}$ r\'epond au probl\`eme de l'acheminement
de messages de mani\`ere instantan\'ement stabilisante
\`a condition que $\mathcal{A}$ s'ex\'ecute de mani\`ere simultan\'ee si
et seulement si :\\
\indent \textbf{(1)} $BG$ est un DAG une fois $\mathcal{A}$ stabilis\'e.\\
\indent \textbf{(2)} $\mathcal{C}$ est \'equitable pour la g\'en\'eration de messages et pour
la transmission de messages.\\
\indent \textbf{(3)} Pour tout message valide $m$ non d\'elivr\'e durant la stabilisation
de $\mathcal{A}$ et quelle que soit l'ex\'ecution de $\mathcal{C}$,
il existe une copie de $m$ situ\'ee dans un buffer ad\'equat pour $m$
une fois $\mathcal{A}$ stabilis\'e.\\
\indent \textbf{(4)} Tous les messages occupant des buffers qui ne leur sont pas ad\'equats
une fois $\mathcal{A}$ stabilis\'e lib\`erent tous les buffers de $FB$
et de $NB$ en un temps fini quelle que soit l'ex\'ecution de $\mathcal{C}$.\\
\indent \textbf{(5)} Pour tout message valide $m$, $\mathcal{C}$ maintient au moins une
copie de $m$ tant que celui-ci n'est pas d\'elivr\'e.
\end{theoreme}
\vspace{-0.25cm}
Le lecteur peut trouver un algorithme r\'epondant \`a ces propri\'et\'es dans \cite{CDV09a}. Nous allons \`a pr\'esent donner les id\'ees de la preuve de ce r\'esultat.

\paragraph{Preuve de la n\'ecessit\'e de la condition}
Nous souhaitons montrer que si $\mathcal{C}$ r\'epond au probl\`eme de l'acheminement
de messages de mani\`ere instantan\'ement stabilisante (\`a condition que
$\mathcal{A}$ s'ex\'ecute de mani\`ere simultan\'ee) alors $\mathcal{C}$
v\'erifie les cinq propri\'et\'es du th\'eor\`eme 2.
Pour cela, nous allons raisonner par contrapos\'ee. Supposons donc que
$\mathcal{C}$ ne v\'erifie pas une des propri\'et\'es du th\'eor\`eme 2.\\
\indent 1) Si $BG$ poss\`ede un circuit une fois $\mathcal{A}$ stabilis\'e, nous pouvons construire un circuit d'attente sur un ensemble de messages et donc un interblocage.\\
\indent 2) Si $\mathcal{C}$ n'est pas \'equitable pour la g\'en\'eration ou la transmission, cela signifie qu'il peut mettre un message en famine et donc l'emp\^echer 
d'\^etre d\'elivr\'e.\\
\indent 3) S'il existe un message valide $m$, non d\'elivr\'e pendant la stabilisation de $\mathcal{A}$, tel qu'aucune copie n'occupe un buffer ad\'equat pour 
$m$ une fois $\mathcal{A}$ stabilis\'e, cela signifie que ce message ne pourra pas \^etre d\'elivr\'e.\\
\indent 4) S'il existe un message ne lib\'erant jamais un buffer de $FB$ ou de $NB$ qui ne lui est pas ad\'equat une fois $\mathcal{A}$ stabilis\'e,
cela signifie qu'il existe un message qui peut \^etre mis en famine pour sa g\'en\'eration (cas de $FB$) ou une de ses transmissions (cas de $NB$).
Dans ce cas, ce message ne pourra pas \^etre d\'elivr\'e.\\
\indent 5) S'il existe un message valide $m$ tel que $\mathcal{C}$ ne maintienne pas une copie tant qu'il n'est pas d\'elivr\'e, cela signifie que $m$ a \'et\'e perdu.
Ce message ne pourra pas \^etre d\'elivr\'e.\\
Dans tous les cas, nous constatons que $\mathcal{C}$ ne r\'epond pas au probl\`eme de l'acheminement de messages 
de mani\`ere instantan\'ement stabilisante, ce qui \'etait le r\'esultat \`a prouver.

\paragraph{Preuve de la suffisance de la condition}
Nous souhaitons montrer que si un contr\^oleur $\mathcal{C}$ v\'erifie les cinq propri\'et\'es
du th\'eor\`eme 2 alors il r\'epond au probl\`eme de l'acheminement
de messages de mani\`ere instantan\'ement stabilisante (\`a condition que
$\mathcal{A}$ s'ex\'ecute de mani\`ere simultan\'ee).

\vspace{-0.35cm}
\begin{lemme}
Si les tables de routage sont correctes dans la configuration initiale
et si $\mathcal{C}$ v\'erifie les cinq propri\'et\'es du th\'eor\`eme 2 alors
il r\'epond au probl\`eme de l'acheminement
de messages de mani\`ere instantan\'ement stabilisante.
\end{lemme}
\vspace{-0.25cm}

Pour prouver ce r\'esultat, il faut reprendre la preuve du th\'eor\`eme 1 (le lecteur pourra se reporter \`a 
la preuve du th\'eor\`eme 5.7 dans \cite{T01}), ce qui n\'ecessite la propri\'et\'e 1 du th\'eor\`eme 2
(en constatant que la propri\'et\'e du th\'eor\`eme 2 4 permet d'assurer que les messages invalides ne perturberont pas la preuve). Cela permet 
de prouver que $\mathcal{C}$ est sans interblocage dans le cas consid\'er\'e. Les propri\'et\'es 2 et 5 du th\'eor\`eme 2 permettent alors de d\'eduire le r\'esultat.

\vspace{-0.35cm}
\begin{lemme}
Si les tables de routage sont quelconques dans la configuration initiale
et si $\mathcal{C}$ v\'erifie les cinq propri\'et\'es du th\'eor\`eme 2 alors
il r\'epond au probl\`eme de l'acheminement
de messages de mani\`ere auto-stabilisante \`a condition que $\mathcal{A}$ s'ex\'ecute simultan\'ement.
\end{lemme}
\vspace{-0.25cm}

Par hypoth\`ese, $\mathcal{A}$ est un algorithme de calcul de table
de routage auto-stabilisant et silencieux. Cela signifie que les tables
de routage seront construites et correctes en un temps fini (gr\^ace
au fait que $\mathcal{A}$ soit prioritaire sur $\mathcal{C}$).
Or, $\mathcal{C}$ s'ex\'ecute de manière simultan\'ee,
l'occupation des buffers va donc \^etre modifi\'ee (acceptation de nouveaux
messages et acheminement de messages).
Nous pouvons appliquer le lemme 1 une fois $\mathcal{A}$ stabilis\'e.
Nous savons donc que $\mathcal{C}$ r\'epond au probl\`eme de l'acheminement
de messages de mani\`ere instantan\'ement stabilisante lorsque les tables de routage
sont correctes et constantes (ce qui est assur\'e par le fait que $\mathcal{A}$
est silencieux). $\mathcal{C}$ r\'epondra donc à la sp\'ecification du
probl\`eme \`a partir de cette configuration. $\mathcal{C}$ est donc
auto-stabilisant.

\vspace{-0.35cm}
\begin{lemme}
Si les tables de routage sont quelconques dans la configuration initiale
et si $\mathcal{C}$ v\'erifie les cinq propri\'et\'es du th\'eor\`eme 2 alors
il ne détruit aucun message valide sans le d\'elivrer.
\end{lemme}
\vspace{-0.25cm}

Il s'agit d'une cons\'equence directe des propri\'et\'es 3 et 5.

Les lemmes 2 et 3 nous permettent de d\'eduire trivialement que
si $\mathcal{C}$ v\'erifie les cinq propri\'et\'es du th\'eor\`eme 2, alors il 
r\'epond au probl\`eme de l'acheminement
de messages de mani\`ere instantan\'ement stabilisante \`a condition que $\mathcal{A}$ s'ex\'ecute simultan\'ement.

\section{Conclusion} 
\label{sec:conclusion}

Dans cet article, nous avons donn\'e un ensemble de propri\'et\'es n\'ecessaires et suffisantes pour obtenir un acheminement de messages
instantan\'ement stabilisant. En ce sens, notre r\'esultat g\'en\'eralise celui obtenu dans \cite{MS78,TS81}. Il serait possible d'appliquer 
notre r\'esultat pour d\'emontrer plus simplement l'algorithme de \cite{CDV09a}.

\bibliographystyle{alpha}
\small{
\bibliography{Biblio}
}
\label{sec:biblio}

\end{document}